\documentclass[a4paper,fleqn,usenatbib]{mnras}
\usepackage{newtxtext,newtxmath}
\usepackage[T1]{fontenc}
\usepackage{ae,aecompl}
\usepackage{float}
\usepackage{color}

\usepackage{graphicx}	
\usepackage{amsmath}	
\usepackage{amssymb}	
\usepackage{xcolor}

\title[Bondi accretion]{Bondi on spherically symmetric accretion}

\author[Philip J. Armitage]{Philip J. Armitage$^{1,2}$
\\
$^{1}$ Department of Physics and Astronomy, Stony Brook University, Stony Brook, NY 11790, USA \\
$^{2}$ Center for Computational Astrophysics, Flatiron Institute, 162 Fifth Avenue, New York, NY 10010, USA \\
}

\begin{document}
\label{firstpage}
\pagerange{\pageref{firstpage}--\pageref{lastpage}}
\maketitle

\begin{abstract}
Hermann Bondi's 1952 paper ``On spherically symmetrical accretion" is recognized as one of the foundations of accretion theory. Although Bondi later remarked that it was ``not much more than an examination exercise", his mathematical analysis of spherical accretion on to a point mass has found broad use across fields of astrophysics that were embryonic or non-existent at the time of the paper's publication. In this non-technical review, I describe the motivations for Bondi's work, and briefly discuss some of the applications of Bondi accretion in high energy astrophysics, galaxy formation, and star formation.
\end{abstract}



A black hole of mass $M$ is immersed in gas, which at large distances is at rest with uniform density $\rho_\infty$ and sound speed $c_\infty$. At what rate does the black hole swallow the gas in its vicinity? A Newtonian version of this apparently simple problem was posed and answered by \citet{bondi52}. If spherical symmetry is maintained, and there is no feedback on to the flow, the accretion rate is,
\begin{equation}
 \dot{M} = 4 \pi \lambda \frac{(GM)^2}{c_\infty^3} \rho_\infty,
\label{eq_Bondi_rate}
\end{equation}
where $G$ is Newton's gravitational constant, and the prefactor $\lambda$ depends upon the adiabatic index of the gas. Bondi's paper with this result is one of the most influential to have been published in {\em Monthly Notices of the Royal Astronomical Society}.
 
Hermann Bondi (1919-2005) was a mathematical physicist, cosmologist, and relativist. He made key contributions to our understanding of the physical nature of gravitational waves \citep{Bondi62}, and was one of the originators of the steady-state cosmological model \citep{bondi48}. Later in life he held a series of high-level posts in public service, including as the second Director General of the European Space Research Organization, a predecessor of today's European Space Agency (ESA). Although he was one of the pre-eminent experts of the day in general relativity, the stimulus for his 1952 paper on accretion had nothing to do with black holes or neutron stars, whose existence and importance for astronomical phenomena would not be recognized for more than another decade. Rather, he was extending a line of research that had been started by Fred Hoyle and Raymond Lyttleton, into the accretion of gas by stars moving through the Interstellar Medium (ISM) at supersonic speeds \citep{hoyle39}. Hoyle and Lyttleton, in turn, were motivated by hypotheses and astronomical problems of their time that may have a quixotic flavour to modern readers. They had suggested that Solar luminosity variations, sourced by changes in the rate of accretion from the ISM, might be the cause of ice ages, and that massive O and B stars could survive for the age of the Galaxy by continually accreting fresh hydrogen fuel. These ideas had stimulated debate \citep{gamow40} but had not found broad favour among the astronomical community. Undeterred by the lukewarm reception to their ideas, a few years later Hoyle suggested that Bondi return to the problem with the goal of putting it ``on a proper mathematical basis"\footnote{Interview of Hermann Bondi by David DeVorkin on March 20 1978, Niels Bohr Library \& Archives, American Institute of Physics, College Park, MD, USA.}. Bondi did just that \citep{bondi44}, and the process by which a gravitating object accretes gas as it moves through a surrounding medium is now known as Bondi-Hoyle-Lyttleton accretion. 

The problem Bondi turned to in 1951 was to calculate the rate of accretion from a uniform medium on to a Newtonian point mass, in the limit where the accreting object is at rest relative to the surrounding gas. Although this is a more symmetric situation than Bondi-Hoyle-Lyttleton accretion, it requires a proper treatment of the hydrodynamics of the inflow, which can be ignored in the simplest description of highly supersonic Bondi-Hoyle-Lyttleton flows. Nonetheless, it did not detain Bondi for long. He solved the problem in the course of just a few days, and later commented that it was a very simple analysis that was ``not much more than an examination exercise" \citep{bondi90}. That description may be overly modest---at least today few instructors would dare to ask for a derivation of Bondi accretion sight unseen---but it is certainly true that the technical difficulties of the calculation would not have fazed mathematical physicists of earlier generations. Bondi's achievement was as much in appreciating that the problem needed solving, as it was in carrying through the calculation. Once he had the solution it was Lyttleton who persuaded him that it was substantial enough to merit publication.

Early citations to \citet{bondi52} focused on whether the model had anything interesting to say about the problems that had motivated Hoyle, Lyttleton and Bondi from the start. It was soon determined beyond doubt that accretion had nothing to do with Solar problems such as the origin of the Sun's corona. As early as 1951, \citet{biermann51} had deduced that the properties of cometary ion tails implied that charged particles were flowing radially {\em outward} from the Sun, and drawing on these observations Eugene Parker developed the first models of the Solar wind \citep{parker58}. The simplest Solar wind models are closely related to the isothermal limit of Bondi accretion, with outflow replacing inflow, and boundary conditions that are specified as $r \rightarrow 0$ rather than $r \rightarrow \infty$. Bondi accretion and the Parker wind are both examples of transonic flows, in which gas accelerates from subsonic to supersonic speeds as it passes through a sonic point. The two problems are mathematically so similar as to be taught jointly in many courses on astrophysical fluid dynamics. 

The role of accretion in the formation and evolution of massive stars took more time, and several observational breakthroughs in infrared and mm-wave astronomy, to elucidate. The original idea that Galactic O and B stars can be almost indefinitely rejuvenated by accretion is wrong, though in a broader context of course these (and all other) stars form as a result of accretion. Strikingly little was known empirically about star formation in the mid-twentieth century. Carbon monoxide, the most important tracer of molecular gas, was not observed astronomically until 1970 \citep{wilson70}, with early surveys of its distribution within the Milky Way following a few years later \citep{scoville75}. These discoveries were steps toward the modern understanding of star formation taking place within molecular clouds, whose density is far higher than  the relatively diffuse phases of the ISM that were known earlier. The physics of the initial collapse of dense molecular gas to form stars necessarily involves the self-gravity of the gas \citep{larson69}, which is not included in Bondi's solution. That said, there are a number of scenarios in which the ``late" accretion of gas by stars (or their discs) within a young stellar cluster could have observable consequences \citep[e.g.][]{bonnell01,throop08,bastian13}. Accretion from a turbulent magnetized medium on to a moving star is a complex problem \citep{burleigh17}, but at heart it is a variation on Bondi-Hoyle-Lyttleton accretion.

Extension of Bondi's work, to include additional physical effects that matter in specific environments, started immediately and continues to this day. The second ever citation to \cite{bondi52} was a prescient paper by \citet{mestel54} entitled ``The influence of stellar radiation on the rate of accretion", that foreshadows what is now a vast literature on the role that feedback plays in accretion processes. Study of the collisionless limit of spherical accretion---which might be appropriate for example for a primordial star interacting with dark matter---started even {\em prior} to work in the fluid approximation \citep{eddington26}, and has been successively extended and improved  \citep{danby57,begelman77}. Analogues of Bondi accretion appropriate for radiation dominated fluids \citep{begelman78}, and for inflows vulnerable to thermal instability and the formation of multiple phases \citep{moscibrodska13}, have been considered and, in more recent years, simulated.

It took the discovery of neutron stars and stellar-mass black holes in X-ray binaries, together with the realization that quasars and other Active Galactic Nuclei are powered by gaseous inflow toward supermassive black holes, to demonstrate the broad importance of accretion as an astrophysical process. Whether in binary systems or galactic nuclei, the surrounding gas invariably has too much angular momentum to accrete spherically. Accretion {\em discs}, whether of the geometrically thin flavour found in luminous sources \citep{lynden-bell69,shakura73,lynden-bell74} or the radiatively inefficient type present in our Galactic Centre \citep{rees82,narayan94,yuan14}, are therefore responsible for most of the observed properties of these systems. Even given this contemporary understanding of the primacy of discs, Bondi's paper remains one of the key foundations of accretion theory. The Bondi accretion rate represents a basic, order of magnitude estimate, for how rapidly gas at large distances can be {\em supplied} to a gravitating object. Another foundation from the same era is the Eddington limit, which represents (also only approximately) the rate at which an accreting object can {\em accept} gas before radiation pressure overwhelms gravity and disrupts the inflow. The interplay between the Bondi and Eddington accretion rates---and the basic fact that the former scales as $M^2$ whereas the latter is linear in $M$---lies at the heart of open questions such as how the first supermassive black holes formed \citep{inayoshi19}.

Many of the recent citations to \citet{bondi52} are in papers using numerical simulations to follow the growth of supermassive black holes in galactic nuclei. The Bondi formula expresses the accretion rate in terms of the density and sound speed of gas at large distances from the accreting object, where ``large" means distances much greater than the {\em Bondi radius},
\begin{equation}
    r_{\rm B} = \frac{2 G M}{c_\infty^2}.
\end{equation}
where the thermal energy of the the accreting gas matches its gravitational potential energy. For the supermassive black hole at the centre of the Milky Way, $r_{\rm B} \simeq 0.05 \ {\rm pc}$, so the Bondi radius is very small compared to galactic scales, which are measured in kpc. The Bondi radius is larger than the scale of the event horizon, however, by a comparably  large factor of the order of $(c/c_\infty)^2 \sim 10^5$ (where $c$ is the speed of light). This scale separation has led to the use of the Bondi formula as a part of sub-grid models for black hole accretion in galactic and cosmological-scale numerical simulations. The need and motivation for these models arises from the fact that black hole accretion releases tremendous amounts of energy, that can feedback to have a profound effect on the structure of the surrounding galaxy or galaxy cluster. Ideally, we would like to simultaneously model both the large-scale cosmological environment of a galaxy, and the small-scale dynamics of black hole accretion within it, but this is computationally impossible. The best that we can do is to resolve down to scales that are of the order of the Bondi radius, and then use a sub-grid model to infer an estimate for how fast the black hole might accrete in that environment.
The fidelity of sub-grid models is hard to assess, and the subject of debate, but they are an indispensable part of numerical simulations of galaxy formation \citep{springel05,booth09,dashyan19}.

Paradoxically, the importance of Bondi's work has  continued to increase in spite of the realization that the Bondi accretion rate is at most rarely, and perhaps never, the correct answer in actual astrophysical systems. The accretion flow on to the supermassive black hole at the centre of the Milky Way furnishes concrete observational evidence. In the Galactic Centre, X-ray observations of the density and temperature of gas at the Bondi radius imply an accretion rate, according to equation~(\ref{eq_Bondi_rate}), that is of the order of $10^{-6} \, M_\odot \, {\rm yr}^{-1}$ \citep{baganoff03}. This is consistent with the rate of mass loss from massive stars in the vicinity, which feed the black hole \citep{cuadra06}. It is not, however, consistent with the inferred accretion rate on the scale of the black hole event horizon, which is hard to pin down precisely but which is estimated to be $\sim 10^{-8} \, M_\odot \, {\rm yr}^{-1}$ \citep{marrone07}. The culprit, once again, is angular momentum. Even in the Galactic Centre, where radiative cooling is inefficient and there is no clear evidence for a geometrically thin accretion disc, the formation of a rotationally supported hot flow means that accretion is accompanied by the release of energy. That energy inevitably feeds back on the inflow and suppresses the small-scale accretion rate \citep{blandford99}. The fact that feedback can suppress  accretion has long been known, but exactly how it works in the context of the Galactic Centre remains the subject of active research \citep{ressler18}. The same is true of the more general question of how and when Bondi-like accretion on large scales can co-exist with non-spherical outflows on smaller scales. At the time of writing, at the end of 2019, the most recent citation to \citet{bondi52} is a paper that addresses this very question \citep{waters20}.

It is amusing to note that the theoretical study of accretion was initiated, in part, by speculations about one of the systems where accretion is {\em not} taking place---the present-day Sun. Despite this false start, accretion is now recognized to be a process that cuts across disparate fields of astrophysics, including star and planet formation, common envelope evolution, X-ray binaries and Active Galactic Nuclei \citep{frank02}. Feedback from supermassive black holes means that the importance of accretion spreads further, into the formation and evolution of galaxies and galaxy clusters. Bondi's contribution to accretion theory, at a time when many of these fields were embryonic or non-existent, was to understand mathematically perhaps its simplest manifestation.

\section*{Acknowledgments}
My thanks to Mitch Begelman and Daniel Proga for their comments on a draft of this essay.

\bibliographystyle{mnras}

\end{document}